# Colossal reversible conductivity switching by room-temperature oxygen-vacancy ordering in Aurivillius oxide films

Song Zhou[1,2,#], Songge Zhang[3,#,*], Lanting Shi[1,#], Ping Zhang[1], Na Li[4], Jiawei Huang[4], Bolin Meng[1], Chuangye Song[1], Shaoxiang Sheng[1], Yang Chai[3], Lede Xian[1], Jinxing Zheng[2,*], Guangyu Zhang[5,*], Kehui Wu[1,*]

[1] *Tsientang Institute for Advanced Study, Zhejiang 310024, China*

[2] *Institute of Plasma Physics, HFIPS, Chinese Academy of Sciences, Hefei 230031, China.*

[3] *Department of Physics and Materials, The Hong Kong Polytechnic University, Hong Kong 999077, China*

[4] *Songshan Lake Materials Laboratory, Dongguan, Guangdong 523808, China*

[5] *Institute of Physics, Chinese Academy of Sciences, Beijing 100190, China*

*Corresponding author

**Email**:

sonzhang@polyu.edu.hk

jxzheng@ipp.ac.cn

gyzhang@iphy.ac.cn

khwu@tias.ac.cn

#These authors contributed equally to this work.

**Abstract**

**Oxygen vacancies are central to the functionality of oxides, yet they typically exist as randomly distributed point defects, limiting the ability to precisely manipulate their collective behavior. Here, we report the room-temperature formation of a long-range-ordered oxygen-vacancy superstructure in single-crystalline Aurivillius-phase $Bi_2WO_6$ thin films via a mild nitrogen-plasma treatment. This structural transformation unlocks a colossal, reversible modulation of electrical conductivity by more than nine orders of magnitude, accompanied by a striking optical transition from transparent to black. Atomic-resolution imaging and spectroscopy reveal that the vacancies selectively order within the perovskite-like $[WO_4]^{2-}$ layers, forming a coherent "defect lattice" that is**

**absent in the pristine film. Oxygen-plasma treatment removes the vacancy superstructure and restores the initial state, whereas subsequent nitrogen-plasma treatment reconstructs it, enabling repeatable room-temperature switching between distinct structural, electronic and optical states. The phenomenon is also observed in another Aurivillius member, $Bi_2MoO_6$, suggesting its generality across the Aurivillius family. These findings establish a new paradigm for atomic-scale defect engineering—using gentle plasma chemistry to construct ordered "defect lattices", opening avenues for reversible property modulation in complex oxides.**

## Introduction

Oxygen vacancies are among the most fundamental and consequential defects in functional oxides. They govern a wide spectrum of phenomena—from oxide ionic conductivity[1,2], electrical conductivity[3,4] and catalytic activity[5,6] to ferroelectricity[7-9], magnetism[10,11], superconductivity[12,13] and metal–insulator transitions[14,15]. Yet, in most oxides, oxygen vacancies exist as randomly distributed point defects, their spatial arrangement dictated by statistical thermodynamics rather than deliberate design. This randomness fundamentally limits the ability to engineer their collective behavior and harness the emergent properties that ordered vacancy arrays could offer.

An ideal scenario would be to organize oxygen vacancies into a long-range-ordered superstructure, where periodic vacancy arrays act as a coherent "defect lattice" that can fundamentally reshape the electronic, optical, and chemical landscape of the host material. However, the controlled construction of a vacancy superstructure, especially in high-quality single-crystalline thin films, has remained a formidable challenge. Conventional approaches such as thermal annealing under reducing atmospheres, ion irradiation, electric-field modulation or chemical deintercalation often suffer from limited uniformity, poor reversibility, or the introduction of additional disorder that obscures the intrinsic effects of vacancy ordering[16-20]. Recently, electric-thermal treatment has achieved vacancy ordering in bulk oxides[21], and oxygen-vacancy ordering has also been reported in oxide thin films under specific processing conditions[9]. However, these approaches require high temperatures or strongly non-equilibrium conditions and are irreversible; room-temperature reversible ordering in high-quality single-crystalline thin films remains elusive. A method that enables mild, reversible, and spatially uniform creation of ordered vacancy structures in complex oxides is yet to be established.

Layered Aurivillius-phase oxides, with their naturally alternating $[Bi_2O_2]^{2+}$ sheets and perovskite-like blocks, offer a particularly attractive platform for addressing this challenge. Their quasi-two-dimensional architecture provides chemically distinct sublattices—the perovskite blocks host the primary charge-transport pathways, while the $[Bi_2O_2]^{2+}$ layers serve as charge reservoirs and structural stabilizers[22-24]. This structural motif naturally leads to the question of whether a gentle chemical stimulus could be used to selectively write and control oxygen vacancy order within specific layers, enabling reversible switching of macroscopic properties.

Here we demonstrate this paradigm in single-crystalline Aurivillius-phase bismuth tungstate, $Bi_2WO_6$ (BWO) thin films. Using an exceptionally mild, room-temperature nitrogen-plasma treatment as a "chemical tweezer"[25,26], we induce the spontaneous formation of a long-range-ordered $V_o$ superstructure confined to the perovskite $[WO_4]^{2-}$ layers. This structural transformation is accompanied by a fully reversible modulation of electrical conductivity by more than nine orders of magnitude and a striking optical transition from transparent to black. Atomic-scale characterizations and spectroscopic analyses directly visualize the periodic ordering of vacancies forming a coherent "defect lattice" that is entirely absent in the pristine film. Transport measurements reveal that this ordered defect state exhibits unconventional electronic behavior. Over a wide temperature range, the resistance follows an empirical power law, $R(T) \propto T^{-0.30}$. At low temperatures, the system crosses over to two-dimensional Mott variable-range hopping, consistent with the layered confinement of the oxygen-vacancy superstructure within the perovskite blocks. Notably, the vacancy superstructure can be erased by oxygen-plasma treatment and fully restored by subsequent nitrogen-plasma cycling, establishing a reversible room-temperature chemical switch that toggles the material between distinct structural, electronic, and optical states. We further observe the same phenomenon in other Aurivillius members $Bi_2MoO_6$, confirming that this is a general feature of the Aurivillius family rather than an isolated case.

## Results and Discussion

### A reversible, nine-order-of magnitude resistivity switch in Aurivillius-phase oxide thin films

The Aurivillius-phase bismuth tungstate, BWO, has a naturally layered architecture comprising alternating $[Bi_2O_2]^{2+}$ sheets and perovskite-like $[WO_4]^{2-}$ slabs[27,28] (Fig. 1a). This layered architecture naturally endows it with a two-dimensional character[29,30] and a high susceptibility to oxygen content modulation[31]. We exploit this susceptibility to demonstrate how a minimal, room-temperature surface intervention can trigger a profound, bulk transformation of the electronic ground state.

Fig. 1b schematically illustrates our approach: a simple, standard laboratory plasma cleaner delivers either nitrogen or oxygen plasma to the film surface at room temperature. The result is visually striking. As-grown epitaxial BWO thin films (grown by pulsed laser deposition) appear nearly transparent to the naked eye (Fig. 1b, lower panel) and are highly insulating, with current-voltage (*I-V*) measurements showing a response below 1 pA at 5 V (Fig. 1c, red line). After exposure to nitrogen plasma, the film turns uniformly black and exhibits a linear, Ohmic *I-V* characteristic passing through the origin (Fig. 1c, blue line). At 5 V, the current reaches the milliampere scale—an increase in conductivity of more than nine orders of magnitude. This magnitude of change places the transition among the most extreme cases of resistivity modulation reported to date[32,33].

The electronic transformation is further corroborated by photoluminescence (PL) spectroscopy (Fig. 1d). The as-grown film shows a pronounced emission peak centered at 467 nm (2.66 eV), consistent with the reported optical bandgap of 2.75 eV[34]. This PL signal is almost completely quenched in the high-conductivity state, reflecting a profound transformation of the overall electronic structure. This electronic transformation occurs without disrupting the host crystalline lattice. X-ray diffraction (XRD) $\theta$–$2\theta$ scans (Fig. 1e) show only the (00l) reflections of phase-pure, *c*-axis-oriented BWO for both insulating and high-conductivity states, with no detectable peak shifts or emergence of secondary phases. Reciprocal space mapping (RSM) around the STO (001) reflections (Fig. 1f) further confirms that epitaxial relationships remain unchanged. These structural probes demonstrate that the resistivity modulation proceeds isostructurally—the lattice itself serves as a pristine template that hosts the electronic transition without being disrupted by it.

Critically, the transition is fully reversible. Switching from nitrogen to oxygen plasma restores both the film's transparency and its insulating *I-V* characteristic (Fig. 1c, red line). Alternating nitrogen- and oxygen-plasma exposures toggles the film between high- and low-conductance states over dozens of cycles with negligible degradation (Fig. 1g), establishing a robust, room-temperature chemical switch that operates with commonplace laboratory equipment. This cyclability, combined with the structural integrity preserved throughout, sets the stage for understanding the atomic-scale origin of the transition. Note that the above experimental observations are preserved for film thickness in the range of 100 nm-1000 nm, with thicker films show stronger signals (darker color, better conductivity after plasma treatment). In all following experiments the film thickness is selected as 500 nm.

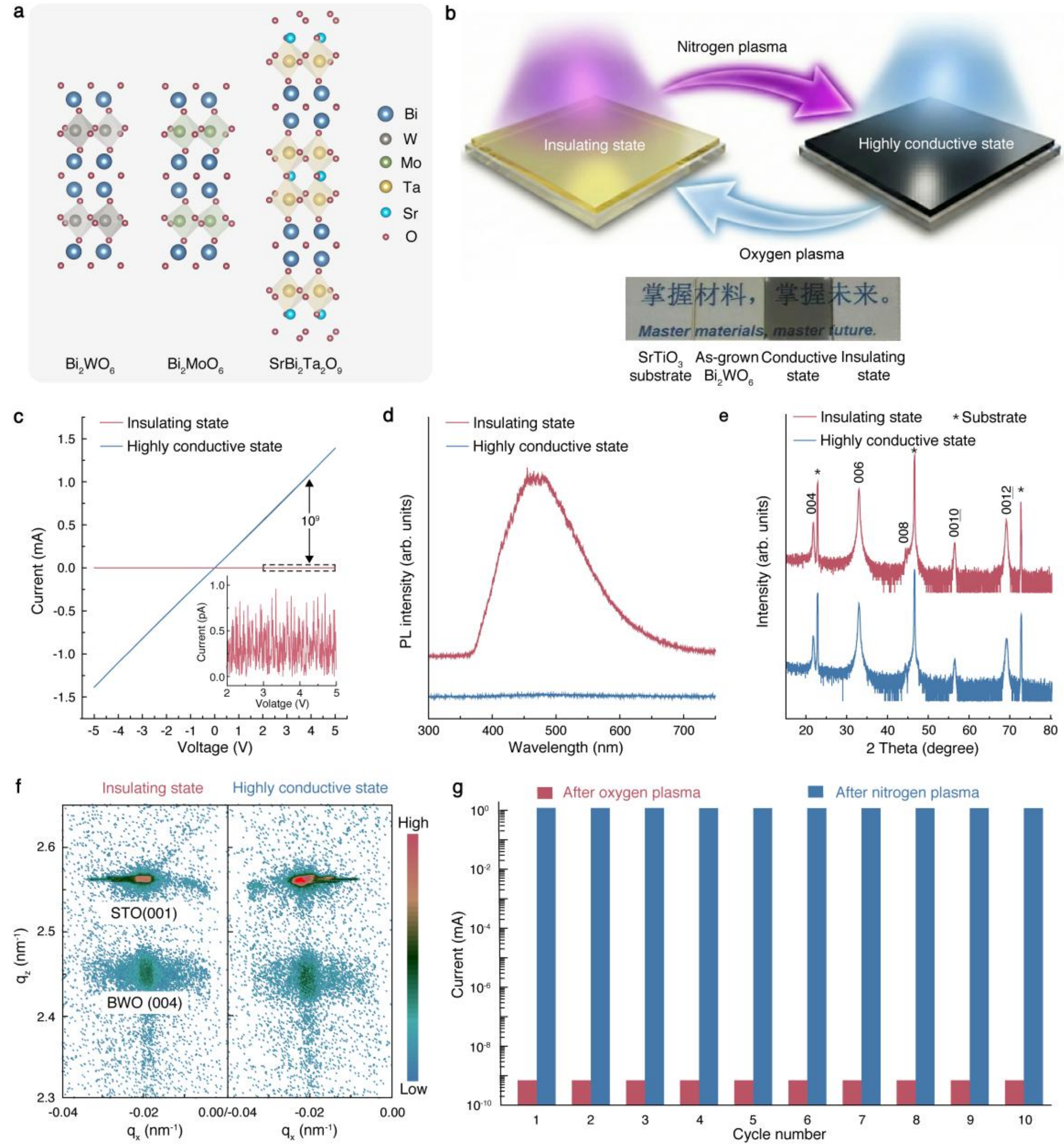

**Figure 1. A colossal, reversible and isostructural resistivity switch induced by surface plasma treatment in Aurivillius-phase $Bi_2WO_6$.** (a) Crystal structures of three representative Aurivillius-phase oxides: $Bi_2WO_6$, $Bi_2MoO_6$, and $SrBi_2Ta_2O_9$, highlighting their common layered motif. (b) Schematic of the surface plasma treatment process (top) and corresponding optical photographs (bottom) showing the reversible visual transition of a BWO thin film between transparent insulating and black conductive states. The film thickness is 150nm. (c) Current–voltage (*I*–*V*) characteristics of a pristine (red, insulating) and a nitrogen-plasma-treated (blue, highly conductive) 150-nm-thick BWO film. The current scale changes by over nine orders of magnitude. (d) Photoluminescence (PL) spectra of the insulating (red) and highly conductive (blue) states. The emission peak at ~467 nm (~2.66 eV) in the insulator is quenched after the plasma treatment. (e) X-ray diffraction (XRD) θ-2θ scans. (f) Reciprocal space maps (RSMs) around the STO (001) reflection. (g) Cyclic switching stability. The device current at a fixed voltage robustly oscillates between conductive (high) and insulating (low) states over multiple

nitrogen- and oxygen-plasma treatment cycles.

### Atomic-scale origin of the reversible resistivity switching

This extreme property change occurs without any detectable alteration in the macroscopic crystal structure (Figs. 1e-f), pointing to a more subtle microscopic origin. To uncover it, we turned to atomically resolved imaging of the crystal structure. Aberration-corrected scanning transmission electron microscopy (STEM) imaging of the film along the [100] direction, which is sensitive primarily to the heavy Bi and W atoms, revealed that the fundamental Aurivillius framework of alternating bismuth-oxygen and tungsten-oxygen layers remains intact after the transition (Supplementary Figs. S1a, b). This result, consistent with our macroscopic data, strongly suggested that the microscopic driver of the reversible resistivity switching lies not in the heavy-element lattice framework but within the oxygen sublattice.

To directly visualize oxygen atoms, we employed integrated differential phase contrast (iDPC)-STEM, a technique capable of simultaneously imaging both heavy and light elements[35]. Figs. 2a-c show iDPC-STEM images of the pristine insulating film viewed along the [100], [010], and $[1\bar{1}0]$ zone axes, respectively. Structural sketches of orthorhombic BWO, positioned to the left of the experimental images, show excellent agreement. Critically, all oxygen atom columns are clearly resolved without noticeable deficiency, confirming the stoichiometric quality of our as-grown films. The characteristic Aurivillius layering of alternating $[Bi_2O_2]^{2+}$ and $[WO_4]^{2-}$ slabs is distinctly visible.

The corresponding images of the highly conductive state, acquired from the same sample after nitrogen-plasma treatment, are presented in Figs. 2d-f. While the overall Aurivillius framework remains intact, a striking and consistent change is observed: the emergence of an ordered pattern of oxygen vacancies. Along the [100] direction (Fig. 2d), the characteristic "armchair" arrangement of two oxygen pairs flanking each W column (seen in Fig. 2a) is replaced by a configuration where only one prominent oxygen column resides on each side. Similarly, along the [010] direction (Fig. 2e), the paired oxygen columns flanking each W site in the pristine structure are replaced by single oxygen columns. This ordered oxygen deficiency is also evident along the $[1\bar{1}0]$ direction (Fig. 2f), where the paired oxygen columns at the interface between the bismuth-oxygen and tungsten-oxygen layers are reduced to single columns. This ordered oxygen-deficient pattern is not a local phenomenon but extends uniformly from the interface to the surface across the entire single-crystalline film (Supplementary Fig. S2). Consistently, in partially transformed films, a sharp phase boundary separating the insulating and highly conductive phases can be directly visualized (Supplementary Fig. S3). While local vacancy ordering is itself challenging to achieve[21,36], our observation of a global, long-range ordered oxygen-vacancy superstructure (OvS) provides the direct atomic-scale cause for the reversible resistivity switching.

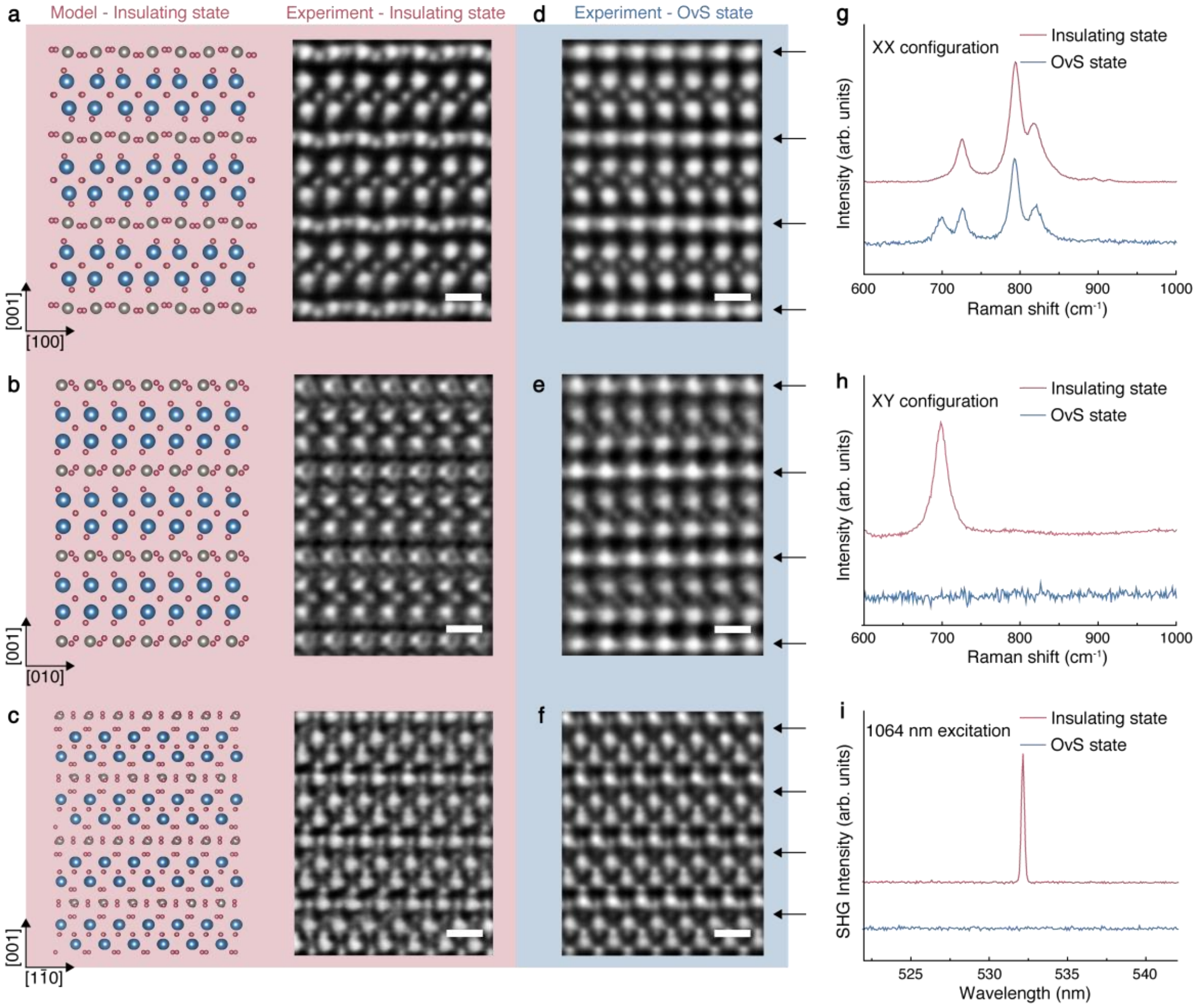

**Figure 2. Atomic-scale origin and symmetry evolution of the reversible resistivity switching.** (a-c) iDPC-STEM images of the pristine insulating $Bi_2WO_6$ film viewed along the [100], [010], and [1$\bar{1}$0] zone axes, respectively. The structural sketches (left panels) show a perfect match. All oxygen atom columns are clearly resolved, confirming a stoichiometric film with intact Aurivillius layering. (d-f) Corresponding iDPC-STEM images of the same sample after nitrogen-plasma treatment, forming the oxygen-vacancy superstructure (OvS) state. While the layered framework persists, a long-range ordered OvS emerges. (g) Raman spectra in the parallel (XX) polarization configuration. (h) Raman spectra in the cross (XY) polarization configuration, all Raman signals in the OvS state are completely quenched, providing definitive evidence for a substantial increase in lattice symmetry, consistent with the ordering effect of the vacancy superstructure. (i) Second-harmonic generation (SHG) signals. The strong SHG response of the ferroelectric insulator is largely suppressed in the OvS state, indicating a global symmetry enhancement. Scale bar :0.5 nm.

This structural transformation is further corroborated by spectroscopic probes. Unpolarized Raman measurements reveal no discernible differences in the main vibrational peaks of BWO within the 600–1000 $cm^{-1}$ range before and after plasma treatment (Supplementary Fig. S4), prompting us to employ polarization-dependent Raman spectroscopy, which is highly sensitive to crystal symmetry and bonding configurations[37]. Figs. 2g and 2h provide decisive evidence for the symmetry evolution. In the parallel (XX) configuration, the OvS state exhibits a distinct peak at 699 $cm^{-1}$ that is absent in the insulator, which instead shows three main modes at 725, 794 and 817 $cm^{-1}$ (Fig. 2g). The cross (XY) configuration, sensitive to asymmetric vibrations, reveals a striking contrast: all Raman signals in the 600–1000 $cm^{-1}$ range are completely suppressed in the highly conductive state, whereas the insulator displays a clear peak at 699 $cm^{-1}$ (Fig. 2h). These observations identify the 699 $cm^{-1}$ mode, which has been assigned to asymmetric stretching of $WO_6$ octahedra involving equatorial oxygen atoms[38,39], as the key spectral signature differentiating the two phases. Its exclusive appearance in the XX configuration for the highly conductive state and in the XY configuration for the insulator signals a fundamental symmetry change of this vibrational mode across the plasma-induced transition. The complete absence of XY-polarized Raman response in the highly conductive state further indicates a substantial increase in lattice symmetry. This symmetry "purification" aligns perfectly with the formation of a long-range ordered OvS, which averages out local anisotropies and restores higher crystalline symmetry while profoundly modifying the electronic structure.

The enhancement of global symmetry is independently confirmed by second-harmonic generation (SHG) and piezoresponse force microscopy (PFM). The strong SHG signal of the pristine ferroelectric BWO is significantly suppressed in the OvS state (Fig. 2i). Concurrently, the well-defined ferroelectric domain structure observed in the pristine film (Supplementary Figs. S5a, b) completely vanishes after the transition (Supplementary Figs. S5c, d). The loss of both SHG response and ferroelectric contrast confirms the loss of inversion-symmetry breaking, consistent with a global symmetry enhancement upon the plasma-induced transition.

**Theoretical modeling and electronic structure of the oxygen-vacancy-superstructure-driven conductive state**

The atomically resolved iDPC-STEM observations provide a definitive structural basis for the highly conductive state. To understand how this ordered defect landscape transforms the electronic ground state, we constructed a first-principles structural model of BWO incorporating the experimentally observed oxygen-vacancy superstructure within the perovskite layers (Figs. 3a-c). Upon full structural relaxation, the system converges to a state with enhanced symmetry: the oxygen atoms adjacent to W sites become coplanar along both the [100] and [010] directions, while the bridging oxygen atoms connecting W and Bi layers along the $[1\bar{1}0]$ direction adopt a centered configuration. These relaxed structural features are in excellent agreement with the atomic-resolution images presented in Figs. 2d–f, validating the fidelity of our theoretical model.

Using this relaxed structure, we calculated the electronic band structure and the partial density of states (PDOS). The pristine insulating phase exhibits a clear bandgap of ~2.9 eV (Fig. 3d), in reasonable agreement with our optical measurements (Fig. 1d). In stark contrast, the oxygen-vacancy-superstructure model yields a fundamentally different electronic landscape: the bandgap closes completely, and the Fermi level cuts through the conduction bands (Fig. 3e, red line). This provides direct electronic-structure evidence for the insulator-to-conductor transition, demonstrating that electron filling of the W 5d conduction manifold drives the system into a highly conductive state. The PDOS at the Fermi level is dominated by W 5d character (~62.5%), with substantial O 2p (~20.6%) and minor Bi 6p (~16.9%) contributions (Supplementary Fig. S6). The dominance of W 5d character indicates that the vacancy-derived states near the Fermi level are primarily located on the W sites. The substantial O 2p contribution originates from the dangling bonds of oxygen atoms adjacent to the vacancy sites, reflecting the local structural distortion induced by the vacancy ordering.

Analysis of the spatial distribution of electrons near the Fermi level (Fig. 3f) reveals that the conduction is predominantly confined within the tungsten–oxygen layers, with minimal weight on the bismuth–oxygen slabs. This preferential localization points to the two-dimensional character of charge transport in the highly conductive state—a feature that could be further tuned by electrostatic gating to potentially induce a two-dimensional electron gas[40].

Together, direct atomic-scale imaging, symmetry-sensitive spectroscopic probes, and first-principles calculations establish that the fully reversible resistivity switching is driven by the formation of a global oxygen-vacancy superstructure within the perovskite layers. This represents a new class of chemically controlled resistivity switching, where minimal surface intervention creates a long-range-ordered defect landscape that fundamentally reconstructs the electronic band structure and modifies the crystal-field symmetry.

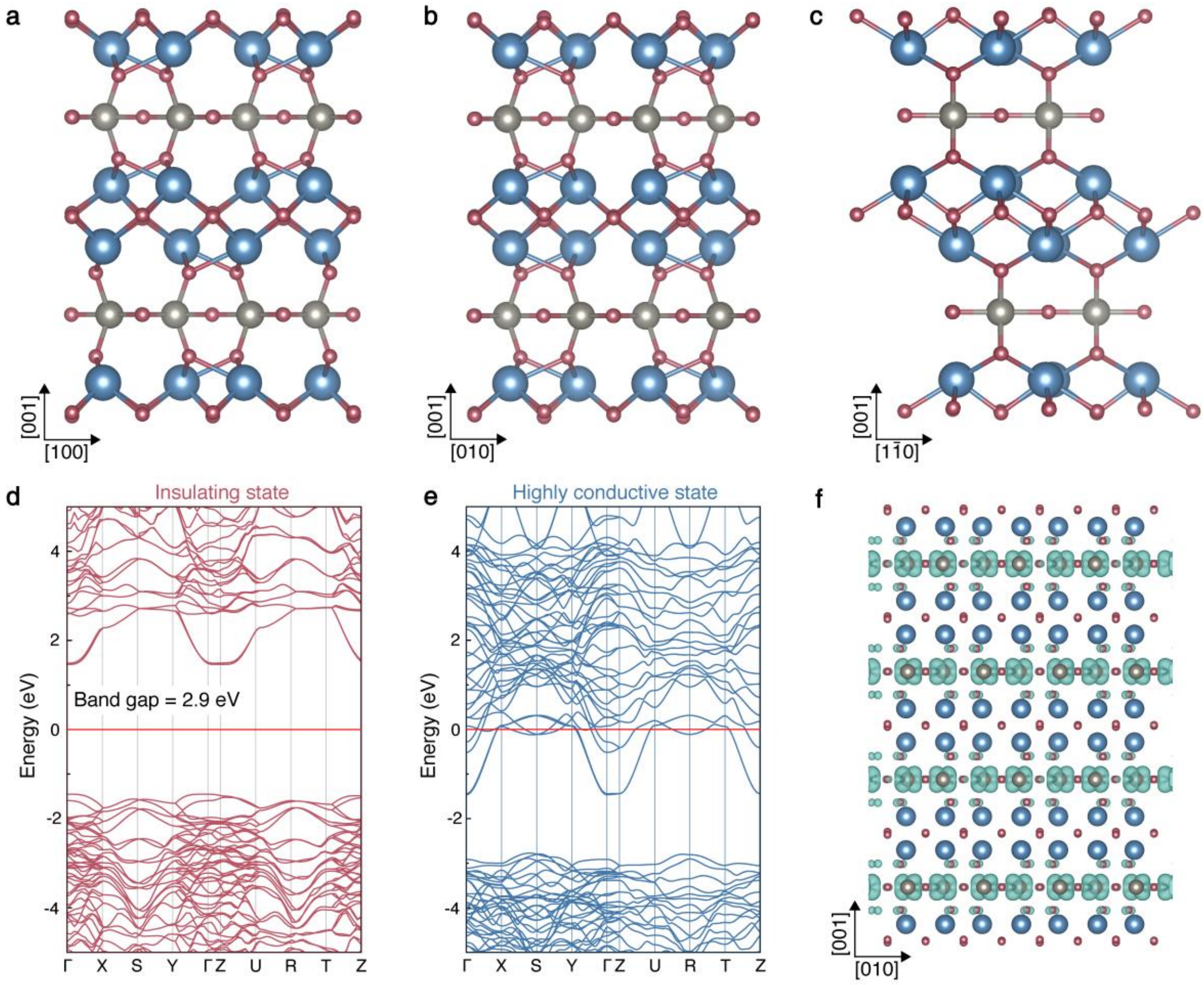

**Figure 3. Theoretical modeling and electronic structure of the oxygen-vacancy-superstructure-driven conductive state**. (a-c) Structural model of highly conductive BWO incorporating the oxygen-vacancy superstructure within the perovskite layers, viewed along the (a) [100], (b) [010], and (c) [1$\bar{1}$0] directions. (d) Calculated electronic band structure of the pristine insulating $Bi_2WO_6$, showing a clear bandgap of ~2.9 eV. (e) Band structure of the oxygen-vacancy-superstructure model, showing complete bandgap closure with the Fermi level (red line) crossing the conduction bands. (f) Spatial distribution of electrons near the Fermi level.

## Transport properties of the highly conductive state

The atomic-scale structural reconstruction and the associated symmetry evolution described above provide a microscopic picture of the oxygen-vacancy-superstructure state. To understand how these microscopic changes manifest in the macroscopic electronic transport properties, we performed electrical and magnetotransport measurements on 500-nm-thick films.

Hall effect measurements were carried out at room temperature using a Hall-bar geometry (Fig. 4a). The Hall resistance $R_{xy}$ exhibits a linear dependence on the magnetic field with a slope of −1.01 Ω/T, indicating electron-type conduction. From the slope, we obtain a carrier concentration of $n = 1.24\times10^{19}$ cm$^{-3}$ and a Hall mobility of $\mu_H = 4.24$ cm$^2$·V$^{-1}$·s$^{-1}$. The measured carrier concentration confirms that the oxygen vacancies introduced by plasma treatment inject a high density of electrons into the conduction band, driving the system into a heavily doped state. The moderate mobility is characteristic of heavily doped transition-metal oxides, where the high carrier density and the accompanying disorder limit the mobility through enhanced ionized-impurity and disorder scattering.

The temperature-dependent resistance $R(T)$ was measured from 300 K down to 1.8 K (Fig. 4b). The resistance increases from 7918 Ω at 300 K to 39261 Ω at 1.8 K, corresponding to resistivity values of $9.90\times10^{-4}$ Ω·m and $4.91\times10^{-3}$ Ω·m, respectively. Over a wide temperature range, the resistance is empirically described by power-law behavior $R(T) \propto T^{-0.30}$ (red solid line in Fig. 4b). This phenomenological relation captures the overall temperature dependence without assuming a specific microscopic mechanism, suggesting that the transport across this broad temperature range may originate from a common underlying physics.

Arrhenius-type thermal activation, commonly observed in conventional semiconductors, fails to describe the low-temperature transport behavior (Supplementary Fig. S7). Instead, at temperatures below 20 K, the system crosses over to a well-defined two-dimensional Mott variable-range hopping (2D VRH) behavior, as evidenced by the linear fit of ln$R$ versus $T^{-1/3}$ (Fig. 4c). This indicates that upon cooling, the system enters a localized regime where conduction proceeds via phonon-assisted hopping between localized states within the two-dimensional tungsten–oxygen layers. The 2D nature of this hopping is consistent with the layered Aurivillius structure and the confinement of the oxygen-vacancy superstructure within the $[WO_4]^{2-}$ layers, as revealed by iDPC-STEM.

X-ray photoelectron spectroscopy (XPS) reveals the electronic consequence of the structural transformation. The pristine insulating film exhibits core-level peaks characteristic of $W^{6+}$ and $Bi^{3+}$ (Figs. 4d, e). After nitrogen plasma treatment, two notable changes occur. First, a low-binding-energy shoulder emerges on the W 4f core level (Fig. 4g), indicating partial reduction of $W^{6+}$ to $W^{5+}$, which provides direct evidence for electron transfer from oxygen vacancies to neighboring W sites[41]. Second, a new component appears in the Bi 4f spectrum at ~162.0 eV and 156.7 eV (Fig. 4f), signaling the reduction of Bi species associated with charge redistribution across the charged $[Bi_2O_2]^{2+}$ layers. This observation reveals that a fraction of the electrons released by oxygen vacancies are redistributed toward $[Bi_2O_2]^{2+}$ layers, driven by the internal electric field inherent to the charged Aurivillius layers, while DFT shows that the itinerant carriers responsible for conduction remain confined to the tungsten–oxygen layers (Fig. 3f). Strikingly, all core levels exhibit a rigid shift of approximately 0.6 eV toward higher binding energy in the highly conductive state. This uniform shift is consistent with the Fermi level moving toward the conduction band, indicative of band-filling effects.

Our DFT calculations provide a consistent microscopic picture that reconciles these observations with the calculated electronic structure. In pristine BWO, the valence band is dominated by O 2p states and the conduction-band minimum by W 5d states, with a computed band gap of 2.9 eV in good agreement with experiment. Upon formation of the ordered oxygen-vacancy superstructure, each vacancy donates two electrons. The partial density of states at the Fermi level is dominated by W 5d character, with substantial O 2p and minor Bi 6p contributions (Supplementary Fig. S6). In the Kohn–Sham limit of a perfectly ordered superstructure, this partially occupied band would render the system conductive.

This expectation is, however, not borne out by the transport measurements. The apparent discrepancy is reconciled as follows: DFT describes the single-particle electronic structure of an idealized, perfectly periodic vacancy arrangement, which does not account for the disorder inherently present in the real film. Spatial fluctuations in the local degree of vacancy ordering, together with point defects, constitute a disorder potential that Anderson-localizes the vast majority of the vacancy-derived states. Consistent with this picture, the Hall carrier density of $1.24\times10^{19}$ $cm^{-3}$ is only ~0.1% of the nominal electron count (~$10^{22}$ $cm^{-3}$ expected from the vacancy concentration), indicating that 99.9% of the electrons introduced by the vacancies remain localized. This picture is further quantified by the localization length extracted from the 2D VRH fit. Using the relation $T_0 = 13.8/[k_B \cdot N(E_F) \cdot \xi^2]$ for 2D Mott VRH[42], with $T_0 \approx 4.1$ K from the VRH fit, we convert the DFT density of states $N(E_F) \approx 10.7$ states $eV^{-1}$ per formula unit to the areal density $N_{2D}(E_F) \approx 1.4$ states $eV^{-1}$ $Å^{-2}$, using the in-plane area of ≈7.4 $Å^2$ per formula unit (a ≈ 5.44 Å, b ≈ 5.46 Å, Z = 4). This yields a localization length $\xi \approx 16.5$ nm, corresponding to approximately 30 in-plane unit-cell lengths. Only a small fraction of the states near the center of the defect band—where the density of states is highest and the localization lengths are longest—remain itinerant and contribute to the measured Hall response; the remaining carriers conduct by hopping, giving rise to the observed variable-range-hopping behavior. The DFT result (conductivity in the ideal ordered limit) and the transport data (localized conduction in the real sample) are therefore not contradictory but complementary: the former establishes that vacancy ordering creates states at the Fermi level with predominantly W 5d character, while the latter demonstrates that these states are largely localized by disorder, yielding a self-consistent picture of a heavily doped oxide with localized carriers.

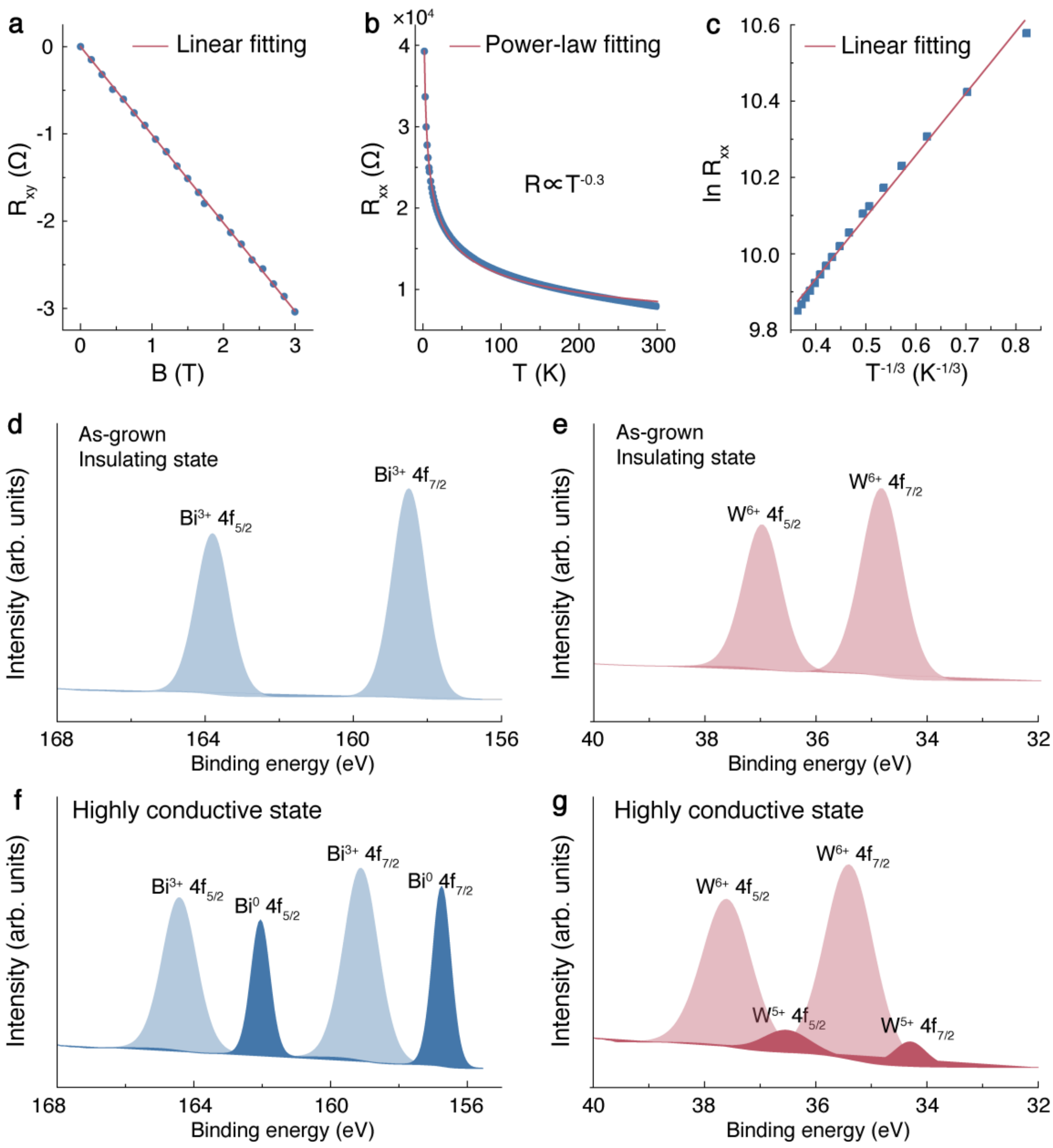

**Figure 4. Transport properties of a 500-nm-thick highly conductive BWO film.** (a) Hall resistance $R_{xy}$ as a function of magnetic field measured at room temperature in a Hall-bar geometry. The linear fit yields a slope of −1.01 Ω/T. (b) Temperature-dependent resistance from 300 K to 1.8 K. The red solid line is a power-law fit $R \propto T^{-0.30}$. (c) Low-temperature resistance plotted as ln$R$ versus $T^{-1/3}$. The linear fit (red solid line) confirms two-dimensional Mott variable-range hopping behavior, consistent with the layered structure of the Aurivillius phase. (d, e) Core-level XPS spectra of the pristine insulating $Bi_2WO_6$ film: (d) Bi 4f and (e) W 4f, showing characteristic $W^{6+}$ and $Bi^{3+}$ peaks. (f, g) Bi 4f and W 4f spectra of the highly conductive film after nitrogen-plasma treatment.

## Generality across the Aurivillius family

To test whether the plasma-induced oxygen-vacancy superstructure formation is unique to BWO or represents a general phenomenon across the Aurivillius family, we extended our study to another prototypical member—$Bi_2MoO_6$ (BMO). Single-crystalline BMO thin films of 100 nm thickness were epitaxially grown on STO (001) substrates (Fig. 5a, left). The high crystalline quality and ferroelectric nature of the as-grown films were confirmed by surface morphology imaging (Supplementary Fig. S8a) and in-plane PFM measurements (Supplementary Fig. S8b, c), which reveal well-defined ferroelectric domains characteristic of single-crystalline BMO.

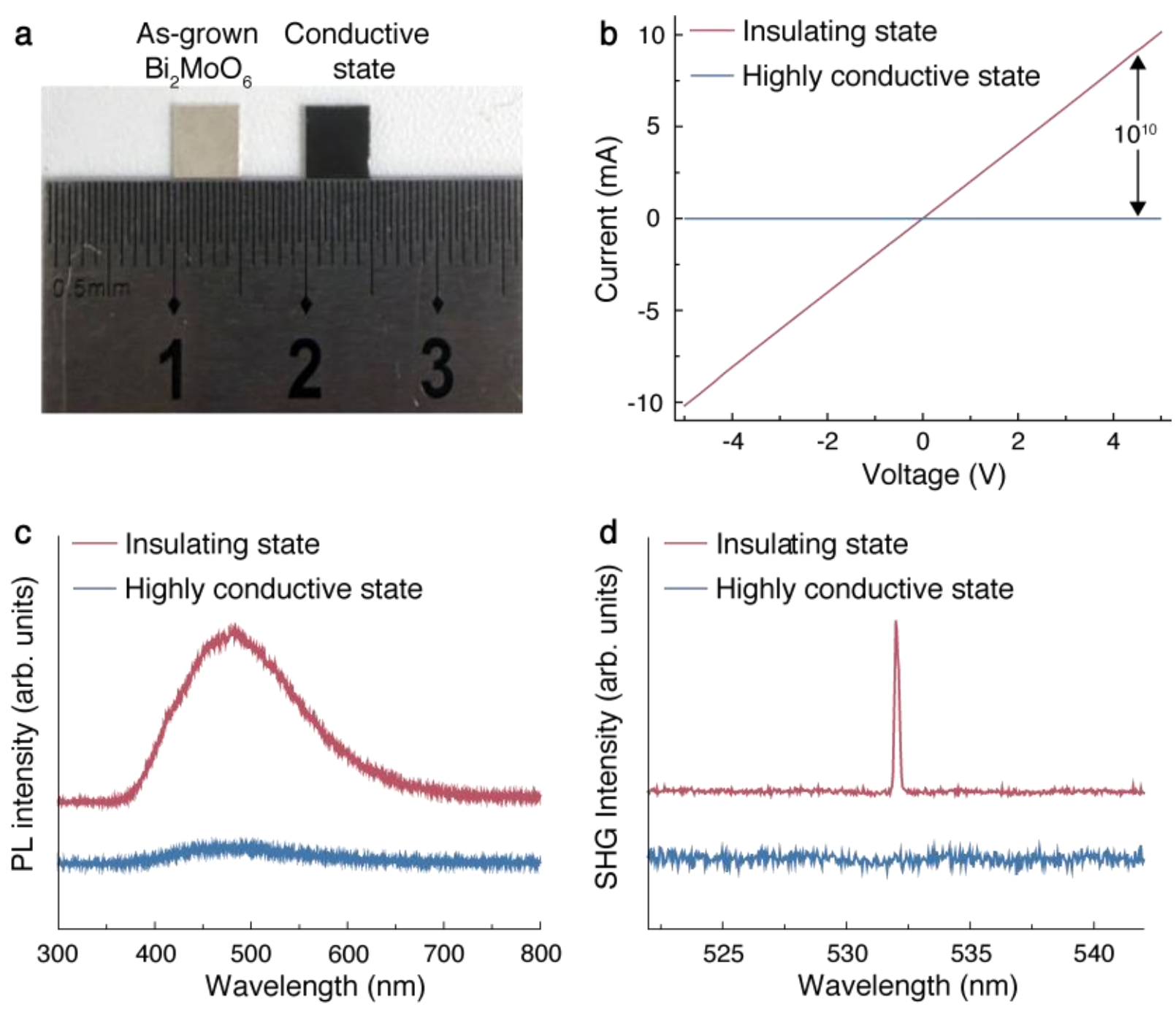

**Figure 5. Generality of the plasma-induced oxygen-vacancy superstructure formation across the Aurivillius family.** (a) Optical image of a pristine 100-nm-thick BMO film grown on STO substrate (left) and the same film after nitrogen-plasma treatment (right), showing the visual transition from transparent to black. (b) I–V characteristics of the pristine insulating BMO film (red) and the nitrogen-plasma-treated highly conductive BMO film (blue), revealing a resistivity modulation exceeding ten orders of magnitude. (c) PL spectra of the pristine insulating BMO film (red), showing an emission peak at ~480 nm, and the highly conductive state (blue), where the PL signal is largely suppressed. (d) SHG signals of the pristine ferroelectric BMO film (red), showing strong response, and the highly conductive state (blue), where the SHG signal is completely quenched, confirming symmetry enhancement.

Upon identical nitrogen-plasma treatment, the BMO film undergoes a striking visual transition from transparent to uniformly black (Fig. 5a, right), mirroring the behavior observed in BWO. Concurrently, the film transforms from an insulating state to a highly conductive state, with a resistivity modulation exceeding ten orders of magnitude (Fig. 5b)—even one order of magnitude larger than the modulation achieved in BWO under the same treatment conditions (Fig. 1c). This enhanced response is consistent with the narrower bandgap of BMO compared to BWO[43], which facilitates a more efficient electronic reconstruction upon oxygen-vacancy ordering.

The electronic transformation is further corroborated by photoluminescence (PL) spectroscopy (Fig. 5c). The pristine BMO film exhibits a distinct emission peak centered at 480 nm (~2.58 eV). After nitrogen-plasma treatment, this PL signal is largely suppressed in the highly conductive state, indicating the formation of an oxygen-vacancy superstructure that effectively modifies the bandgap. Moreover, second-harmonic generation (SHG) measurements reveal that the strong ferroelectric SHG response of the pristine BMO is entirely suppressed in the highly conductive state (Fig. 5d), confirming a substantial increase in structural symmetry—consistent with the symmetry "purification" observed in BWO. These combined results demonstrate that the plasma-induced formation of oxygen-vacancy superstructures and the associated reversible resistivity switching are not unique to BWO but represent a general feature of the Aurivillius oxide family. They highlight the broad applicability of this mild, room-temperature defect-engineering approach.

The same reversible resistivity switching and optical transition were also observed when nitrogen plasma was replaced by hydrogen ($H_2$) or ammonia ($NH_3$) plasma under otherwise identical conditions. The visual color change is nearly indistinguishable across all three plasma species (Supplementary Fig. S9). This consistency strongly indicates that the key mechanism is the reductive nature of the plasma environment—which promotes oxygen vacancy formation and ordering—rather than the incorporation of specific dopant species (e.g., nitrogen). The observation that multiple reducing plasmas produce the same effect further underscores the robustness and generality of this defect-engineering approach, and suggests its potential for integration into diverse plasma-based processing platforms.

**Mechanism of the plasma-induced bulk vacancy ordering**

While the iDPC-STEM measurements unambiguously establish that the oxygen-vacancy superstructure extends uniformly across the entire film thickness from the free surface down to the substrate interface, it is instructive to comment on the underlying kinetics under room-temperature conditions. Given the negligible thermally-activated oxygen bulk diffusion in complex oxides at room temperature, conventional thermally-driven oxygen migration cannot account for the global formation of OvS throughout the 500 nm thick film within only three-minute plasma exposure.

We rationalize this counter-intuitive observation as a cooperative defect-rearrangement process triggered by the near-surface redox perturbation imposed by nitrogen plasma. The reducing plasma environment preferentially removes oxygen from the near-surface region, generating an initial population of oxygen vacancies. Because the oxygen-vacancy formation energy decreases with rising Fermi level (Fig.3/S6), electron accumulation lowers the formation energy of vacancies throughout the film, not merely at the surface. Driven by local chemical-potential gradients and strong inter-defect correlations within the layered Aurivillius lattice, these vacancies undergo collective, cooperative structural reconstruction rather than random thermal diffusion. This defect-mediated cooperative rearrangement propagates inward across the perovskite-like $[WO_4]^{2-}$ slabs, giving rise to the long-range-ordered vacancy superstructure throughout the whole film, without destroying the host Aurivillius framework. This picture is further supported by our observation of atomically-sharp phase boundaries between insulating and OvS domains in partially-converted films (Supplementary Fig.S3), which is characteristic of a collective structural transition instead of gradual vacancy-diffusion-driven penetration. We stress that this process is distinct from high-temperature thermal reduction, which usually produces disordered vacancy distributions. The reversibility of the transition under $O_2$ plasma — which restores oxygen and depletes electrons — corroborates this electronically driven picture.

## Conclusion

In summary, we have demonstrated a chemistry-mediated pathway to achieve a colossal and fully reversible resistivity switching in single-crystalline Aurivillius-phase oxides. Using a simple room-temperature nitrogen-plasma treatment, we induce a conductivity jump of more than nine orders of magnitude, transforming a transparent insulating film into a black conductive film. This macroscopic switching originates from the spontaneous formation of a long-range-ordered oxygen-vacancy superstructure within the perovskite layers—a coherent "defect lattice" that forms without disrupting the host crystal framework. Transport measurements further reveal that this ordered defect state exhibits localized electronic behavior, including non-integer power-law resistivity and low-dimensional variable-range hopping, underscoring the rich physics emerging from the vacancy superstructure.

Our work establishes a new paradigm for atomic-scale defect engineering. The OvS state is not a metastable artifact; it is a robust, cyclable phase that can be repeatedly toggled between distinct structural, electronic, and optical configurations. The generality of this phenomenon across multiple Aurivillius members suggests that this is an intrinsic feature of this oxide family rather than an isolated case, opening the door to a broader class of layered oxides where ordered vacancy architectures can be harnessed for property modulation.

Beyond the immediate demonstration of this reversible resistivity switching, this work offers a versatile platform for exploring emergent phenomena in ordered defect systems. The ability to precisely control vacancy order at the atomic scale, combined with the richness of the Aurivillius structure, provides opportunities for designing novel phases with tailored electronic, optical, and potentially magnetic functionalities. The reversible, room-temperature operation, coupled with the simplicity of the plasma-based method, positions this approach for practical integration into next-generation electronic and photonic devices.

## Methods

### Thin film growth

$Bi_2WO_6$ and $Bi_2MoO_6$ thin films were epitaxially grown on (001)-oriented $SrTiO_3$ or (110)-oriented $NdGaO_3$ substrates by pulsed laser deposition using a 248 nm KrF excimer laser. For $Bi_2WO_6$, the deposition was carried out at a substrate temperature of 740°C under an oxygen pressure of 35 Pa, with a laser energy density of approximately 1.5 J $cm^{-2}$ and a repetition rate of 2 Hz. For $Bi_2MoO_6$, the growth conditions were optimized at a substrate temperature of 650°C and an oxygen pressure of 47 Pa, using a laser energy density of 1.0 J $cm^{-2}$ and the same repetition rate of 2 Hz. After deposition, all samples were cooled to room temperature under an oxygen atmosphere of 2000 Pa at a cooling rate of 5°C $min^{-1}$.

### Plasma treatment

Nitrogen and oxygen plasma treatments were performed using a standard laboratory plasma cleaner. A dedicated vacuum system equipped with a plasma generator was used for the $H_2$ and $NH_3$ plasma treatments. The nitrogen plasma treatment was carried out at a power of 30 W and a pressure of 300 Pa, while the oxygen plasma treatment was conducted at 100 W and 200 Pa. All treatments were performed at room temperature. The duration of each treatment was 3 minutes.

### Piezoresponse force microscopy (PFM)

PFM measurements were performed using an Oxford Instruments Cypher S atomic force microscope, equipped with conductive Pt/Ir-coated silicon probes. The local electromechanical response was characterized in both out-of-plane and in-plane directions using AC excitation amplitudes of 800 mV at frequencies of 350 kHz and 650 kHz, respectively.

**Scanning transmission electron microscopy (STEM)**

Cross-sectional and planar-view TEM specimens for both TEM and STEM observations were prepared through a conventional procedure of sample preparation involving slicing, bonding, mechanical polishing, dimpling, and final ion milling. The HAADF-STEM and iDPC-STEM images were acquired using a ThermoFisher Spectra 300 X-FEG aberration-corrected STEM equipped with dual Cs correctors and a monochromator, operated at 300 kV.

**Materials characterization**

X-ray θ–2θ scan and reciprocal space mappings were performed using a high-resolution X-ray diffractometer (Bruker, D8 Advance). Photoluminescence spectroscopy, second-harmonic generation and Raman spectroscopy was performed on a Horiba LabRam HR Evolution system. X-ray photoelectron spectroscopy (XPS) measurements were conducted on a Thermo Fisher ESCALAB XI+ spectrometer using a monochromatic Al Kα X-ray source. All binding energies were calibrated relative to the adventitious carbon C 1*s* peak at 284.8 eV.

**Electrical and magnetotransport measurements**

Characterization of electrical properties was performed using a probe station (Janis ST-500-1-(4CX)) equipped with Agilent semiconductor analyzers (models 4156C and B1500A). Magnetotransport measurements were conducted in a Physical Property Measurement System (PPMS, Quantum Design) employing standard six-probe Hall-bar configurations.

**Computational details**

All first-principles calculations were carried out using the Vienna Ab initio Simulation Package (VASP) within the framework of density functional theory (DFT). The electron-ion interactions were described by the projector augmented-wave (PAW) method, and the exchange-correlation functional was treated within the generalized gradient approximation (GGA) using the Perdew-Burke-Ernzerhof (PBE) functional. A plane-wave cutoff energy of 500 eV was used. The Brillouin zone was sampled with a Γ-centered 8 × 8 × 3 k-point mesh. Van der Waals interactions were taken into account by including the DFT-D3 correction. The electronic self-consistent field calculations were converged to $10^{-6}$ eV, and ionic relaxation were performed until the residual forces on all atoms were smaller than 0.01 eV/Å. During the structural optimization, both the lattice parameters and the atomic positions were fully relaxed. The partial charge density of $Bi_2WO_{5.5}$ was calculated by integrating the electronic states within an energy window of ± 0.1 eV around the Fermi level.

**Acknowledgements**

This work was supported by Hangzhou Tsientang Education Foundation.

**Author contributions**

K. Wu conceived and designed the project. S. Zhou prepared the thin-film materials and performed structural and physical property characterizations, as well as part of the transport measurements. S. Zhang performed the nitrogen-plasma modification of the materials, device fabrication, and transport measurements, carried out in the group of G. Zhang and under his supervision. L. Shi performed the first-principles calculations under the supervision of L. Xian. S. Zhou and K. Wu prepared the manuscript with the input from all authors. All authors participated in discussions of the project, and made substantial contributions to this work.

**Competing Interests Statement**

The authors declare no competing interests.

**Reference**

1 Yashima, M. *et al.* High oxide-ion conductivity through the interstitial oxygen site in $Ba_7Nb_4MoO_{20}$-based hexagonal perovskite related oxides. *Nature Communications* **12**, 556 (2021). https://doi.org:10.1038/s41467-020-20859-w

2 Pang, S. *et al.* Disordered vacancy-isolated Ce-Gd-O clusters achieve exceptional low-temperature oxygen-ion conductivity for fuel cells. *Science Advances* **12**, eaec8053 (2026). https://doi.org:10.1126/sciadv.aec8053

3 Veal, B. W. *et al.* Interfacial control of oxygen vacancy doping and electrical conduction in thin film oxide heterostructures. *Nature Communications* **7**, 11892 (2016). https://doi.org:10.1038/ncomms11892

4 Evans, D. M. *et al.* Conductivity control via minimally invasive anti-Frenkel defects in a functional oxide. *Nature Materials* **19**, 1195-1200 (2020). https://doi.org:10.1038/s41563-020-0765-x

5 Li, Z. *et al.* Prediction of perovskite oxygen vacancies for oxygen electrocatalysis at different temperatures. *Nature Communications* **15**, 9318 (2024). https://doi.org:10.1038/s41467-024-53578-7

6 Lee, G. R. *et al.* Unraveling oxygen vacancy-driven catalytic selectivity and hot electron generation on heterointerfaces using nanostructured platform. *Nature Communications* **16**,

2909 (2025). https://doi.org:10.1038/s41467-025-57946-9
7 Li, H. *et al.* Role of oxygen vacancies in colossal polarization in $SmFeO_{3-\delta}$ thin films. *Science Advances* **8**, eabm8550 (2022). https://doi.org:10.1126/sciadv.abm8550
8 Wang, L. *et al.* Oxygen Vacancies Generated Out-of-Plane and In-Plane Ferroelectricity in Layered $Bi_2WO_6$ Nanoflake. *Advanced Functional Materials* **35**, 2419399 (2025). https://doi.org/10.1002/adfm.202419399
9 Yang, M. *et al.* Oxygen-Vacancies-Ordering Triggered Large Ferroelectric Polarization in $CaTiO_3$ Thin Films. *Journal of the American Chemical Society* **147**, 21068-21076 (2025). https://doi.org:10.1021/jacs.5c06244
10 Salluzzo, M. *et al.* Origin of Interface Magnetism in $BiMnO_3/SrTiO_3$ and $LaAlO_3/SrTiO_3$ Heterostructures. *Physical Review Letters* **111**, 087204 (2013). https://doi.org:10.1103/PhysRevLett.111.087204
11 Lin, C. & Demkov, A. A. Consequences of Oxygen-Vacancy Correlations at the $SrTiO_3$ Interface. *Physical Review Letters* **113**, 157602 (2014). https://doi.org:10.1103/PhysRevLett.113.157602
12 Marinković, S. *et al.* Direct Visualization of Current-Stimulated Oxygen Migration in $YBa_2Cu_3O_{7-\delta}$ Thin Films. *ACS Nano* **14**, 11765-11774 (2020). https://doi.org:10.1021/acsnano.0c04492
13 Dong, Z. *et al.* Visualization of oxygen vacancies and self-doped ligand holes in $La_3Ni_2O_{7-\delta}$. *Nature* **630**, 847-852 (2024). https://doi.org:10.1038/s41586-024-07482-1
14 Jeong, J. *et al.* Suppression of Metal-Insulator Transition in $VO_2$ by Electric Field–Induced Oxygen Vacancy Formation. *Science* **339**, 1402-1405 (2013). https://doi.org:10.1126/science.1230512
15 Jeen, H. *et al.* Reversible redox reactions in an epitaxially stabilized $SrCoO_x$ oxygen sponge. *Nature Materials* **12**, 1057-1063 (2013). https://doi.org:10.1038/nmat3736
16 Lee, J. *et al.* Selective reduction in epitaxial $SrFe_{0.5}Co_{0.5}O_{2.5}$ and its reversibility. *Nature Communications* **16**, 7391 (2025). https://doi.org:10.1038/s41467-025-62612-1
17 Zhong, H. *et al.* Ion Irradiation Inducing Oxygen Vacancy-Rich $NiO/NiFe_2O_4$ Heterostructure for Enhanced Electrocatalytic Water Splitting. *Small* **17**, 2103501 (2021). https://doi.org/10.1002/smll.202103501
18 Zhang, Q. *et al.* Atomic-resolution imaging of electrically induced oxygen vacancy migration and phase transformation in $SrCoO_{2.5-\sigma}$. *Nature Communications* **8**, 104 (2017). https://doi.org:10.1038/s41467-017-00121-6
19 Das, S. *et al.* Controlled manipulation of oxygen vacancies using nanoscale flexoelectricity. *Nature Communications* **8**, 615 (2017). https://doi.org:10.1038/s41467-017-00710-5
20 Lu, N. *et al.* Electric-field control of tri-state phase transformation with a selective dual-ion switch. *Nature* **546**, 124-128 (2017). https://doi.org:10.1038/nature22389

21 Chen, K. *et al.* A facile approach for generating ordered oxygen vacancies in metal oxides. *Nature Materials* **24**, 835-842 (2025). https://doi.org:10.1038/s41563-025-02171-4

22 Yang, Y. *et al.* Large Switchable Photoconduction within 2D Potential Well of a Layered Ferroelectric Heterostructure. *Advanced Materials* **32**, 2003033 (2020). https://doi.org/10.1002/adma.202003033

23 Song, C. *et al.* Atomic-Scale Characterization of Negative Differential Resistance in Ferroelectric $Bi_2WO_6$. *Advanced Functional Materials* **32**, 2105256 (2022). https://doi.org/10.1002/adfm.202105256

24 Huo, C. *et al.* Low-temperature oxide-ion conduction in Aurivillius-type $((Na_{0.5}Bi_{0.5})_{n-1}Ti_nO_{3n})(Bi_2O_2)$ phases. *Nature Energy* (2026). https://doi.org:10.1038/s41560-026-02115-5

25 Fan, R. *et al.* Plasma surface engineering for efficient and stable perovskite solar cells and modules. *Science* **393**, 490-497 (2026). https://doi.org:10.1126/science.aeg1730

26 Lee, I. *et al.* Barrier-Assisted Plasma Doping for Spatially Selective Resistance Engineering in $MoS_2$ Transistors. *Advanced Materials* **n/a**, e74389 (2026). https://doi.org/10.1002/adma.74389

27 Wang, C. *et al.* Ferroelastic switching in a layered-perovskite thin film. *Nature Communications* **7**, 10636 (2016). https://doi.org:10.1038/ncomms10636

28 Zhou, S. *et al.* Templated perpendicular ferroelectricity in textured Aurivillius oxide-based thin films. *Nature Communications* **17**, 3890 (2026). https://doi.org:10.1038/s41467-026-70676-w

29 Zhou, Y. *et al.* Monolayered $Bi_2WO_6$ nanosheets mimicking heterojunction interface with open surfaces for photocatalysis. *Nature Communications* **6**, 8340 (2015). https://doi.org:10.1038/ncomms9340

30 Zhou, S. *et al.* Ferroelectricity in Epitaxial Perovskite Oxide $Bi_2WO_6$ Films with One-Unit-Cell Thickness. *Nano Letters* **23**, 7838-7844 (2023). https://doi.org:10.1021/acs.nanolett.3c01426

31 Ferenc Segedin, D. *et al.* Topochemical Oxidation of Ruddlesden–Popper Nickelates Reveals Distinct Structural Family: Oxygen-Intercalated Layered Perovskites. *Journal of the American Chemical Society* **148**, 5873-5880 (2026). https://doi.org:10.1021/jacs.5c12712

32 Zhang, G. D. *et al.* Tunable metal-insulator transition in strained $V_2O_3$ thin films epitaxially grown on SiC substrates. *Physical Review Materials* **8**, 035001 (2024). https://doi.org:10.1103/PhysRevMaterials.8.035001

33 Hong, H. *et al.* Metal-to-insulator transition in oxide semimetals by anion doping. *Interdisciplinary Materials* **3**, 358-368 (2024). https://doi.org/10.1002/idm2.12158

34 Liu, X., Long, P., Sun, Z. & Yi, Z. Optical, electrical and photoelectric properties of layered-perovskite ferroelectric $Bi_2WO_6$ crystals. *Journal of Materials Chemistry C* **4**, 7563-7570 (2016). https://doi.org:10.1039/C6TC02069K

35 Lazić, I., Bosch, E. G. T. & Lazar, S. Phase contrast STEM for thin samples: Integrated differential phase contrast. *Ultramicroscopy* **160**, 265-280 (2016).

https://doi.org/10.1016/j.ultramic.2015.10.011

36 Zhang, Q. *et al.* Near-room temperature ferromagnetic insulating state in highly distorted $LaCoO_{2.5}$ with $CoO_5$ square pyramids. *Nature Communications* **12**, 1853 (2021). https://doi.org:10.1038/s41467-021-22099-y

37 Zhang, X., Tan, Q.-H., Wu, J.-B., Shi, W. & Tan, P.-H. Review on the Raman spectroscopy of different types of layered materials. *Nanoscale* **8**, 6435-6450 (2016). https://doi.org:10.1039/C5NR07205K

38 Maczka, M. *et al.* Phonons in ferroelectric $Bi_2WO_6$: Raman and infrared spectra and lattice dynamics. *Applied Physics Letters* **92**, 112911 (2008). https://doi.org:10.1063/1.2896312

39 Djani, H., Hermet, P. & Ghosez, P. First-Principles Characterization of the $P2_1ab$ Ferroelectric Phase of Aurivillius $Bi_2WO_6$. *The Journal of Physical Chemistry C* **118**, 13514-13524 (2014). https://doi.org:10.1021/jp504674k

40 Djani, H. *et al.* Rationalizing and engineering Rashba spin-splitting in ferroelectric oxides. *npj Quantum Materials* **4**, 51 (2019). https://doi.org:10.1038/s41535-019-0190-z

41 Corby, S. *et al.* Water Oxidation and Electron Extraction Kinetics in Nanostructured Tungsten Trioxide Photoanodes. *Journal of the American Chemical Society* **140**, 16168-16177 (2018). https://doi.org:10.1021/jacs.8b08852

42 Ortuño, M., Estellés-Duart, F. & Somoza, A. M. Numerical Simulations of Variable-Range Hopping. *physica status solidi (b)* **259**, 2100340 (2022). https://doi.org/10.1002/pssb.202100340

43 Liu, X., Gu, S., Zhao, Y., Zhou, G. & Li, W. $BiVO_4$, $Bi_2WO_6$ and $Bi_2MoO_6$ photocatalysis: A brief review. *Journal of Materials Science & Technology* **56**, 45-68 (2020). https://doi.org/10.1016/j.jmst.2020.04.023

# Supplementary information

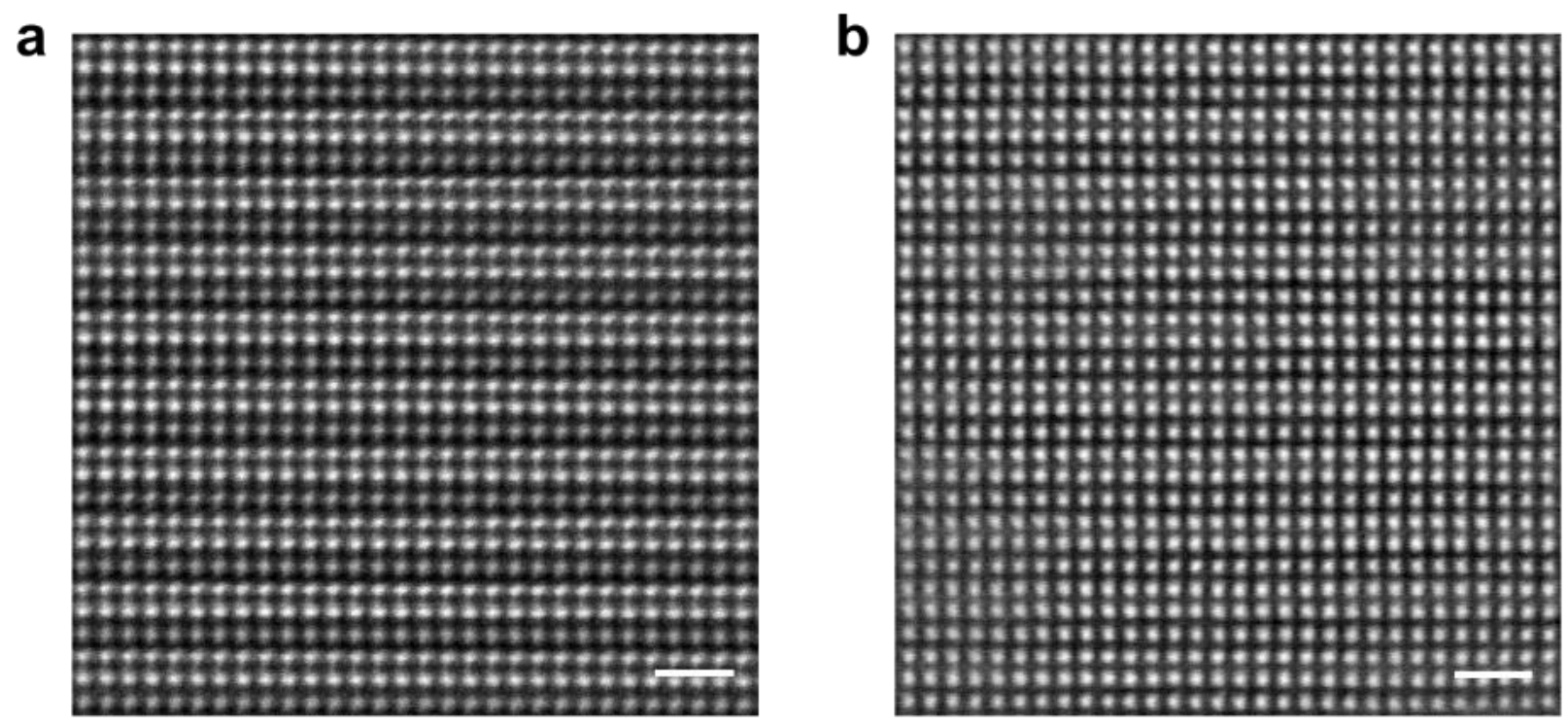

**Figure S1.** HAADF-STEM images of the (a) pristine insulating film and the (b) nitrogen-plasma treated film along the in-plane [100] direction. Scale bar: 1 nm.

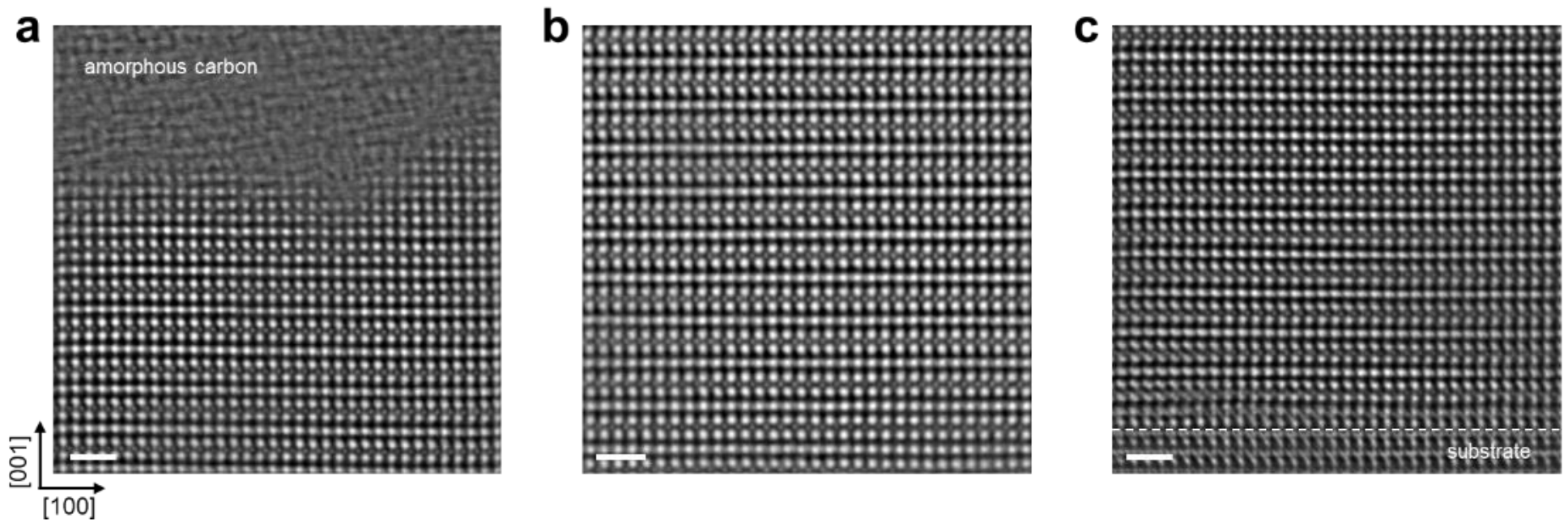

**Figure S2. Global oxygen-vacancy superstructure in the nitrogen-plasma treated film.** iDPC-STEM images of the highly conductive Aurivillius-phase film acquired (a) near the surface, (b) in the middle of the film, and (c) near the interface. The long-range ordered oxygen-vacancy superstructure is observed uniformly across all regions, confirming its global nature throughout the entire single-crystalline film. Scale bar: 1 nm.

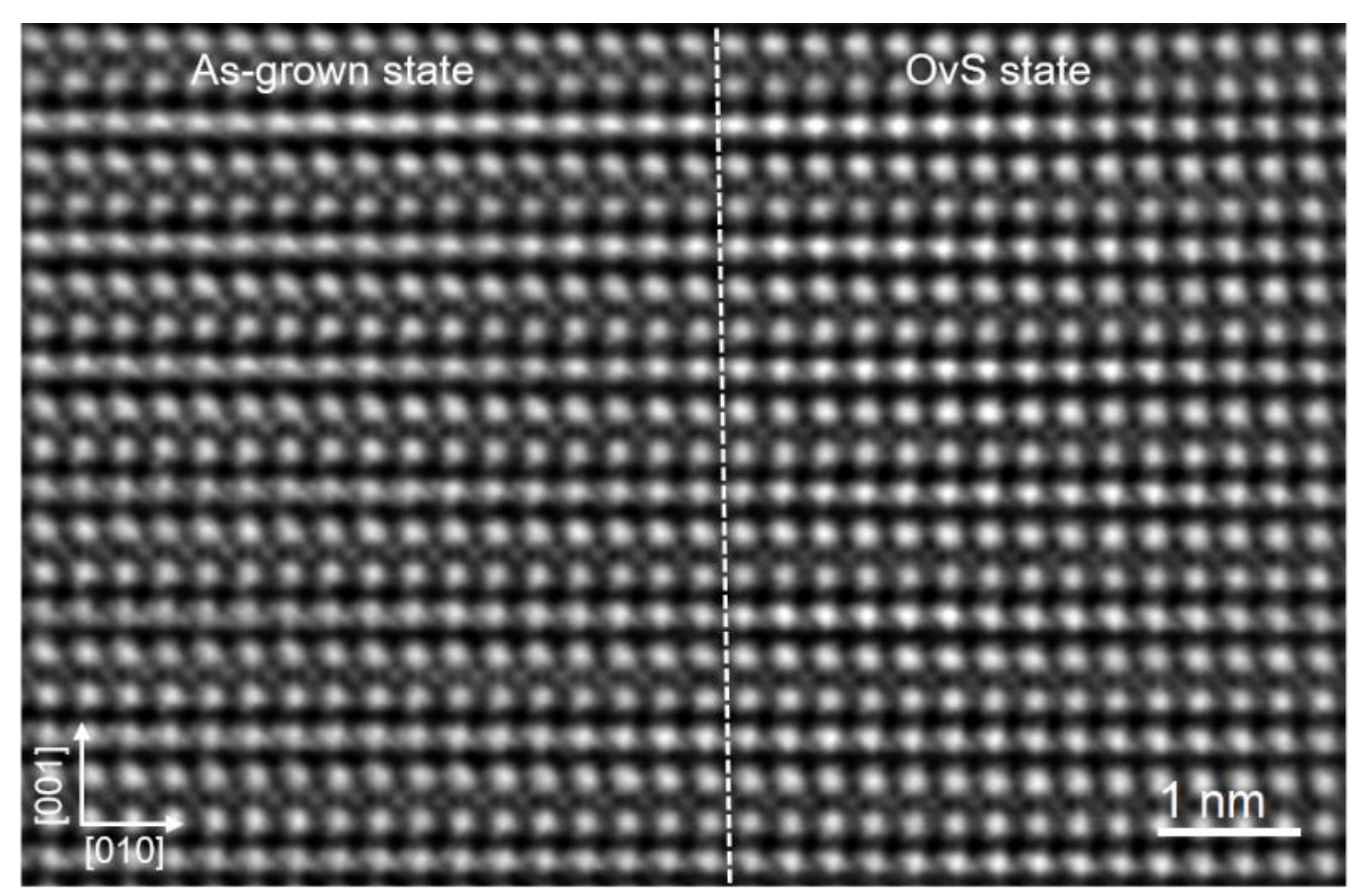

**Figure S3.** Atomic-scale visualization of the phase boundary in an incompletely transformed film. iDPC-STEM image showing the transition from the insulating phase to the highly conductive phase. The dashed white line indicates the phase boundary.

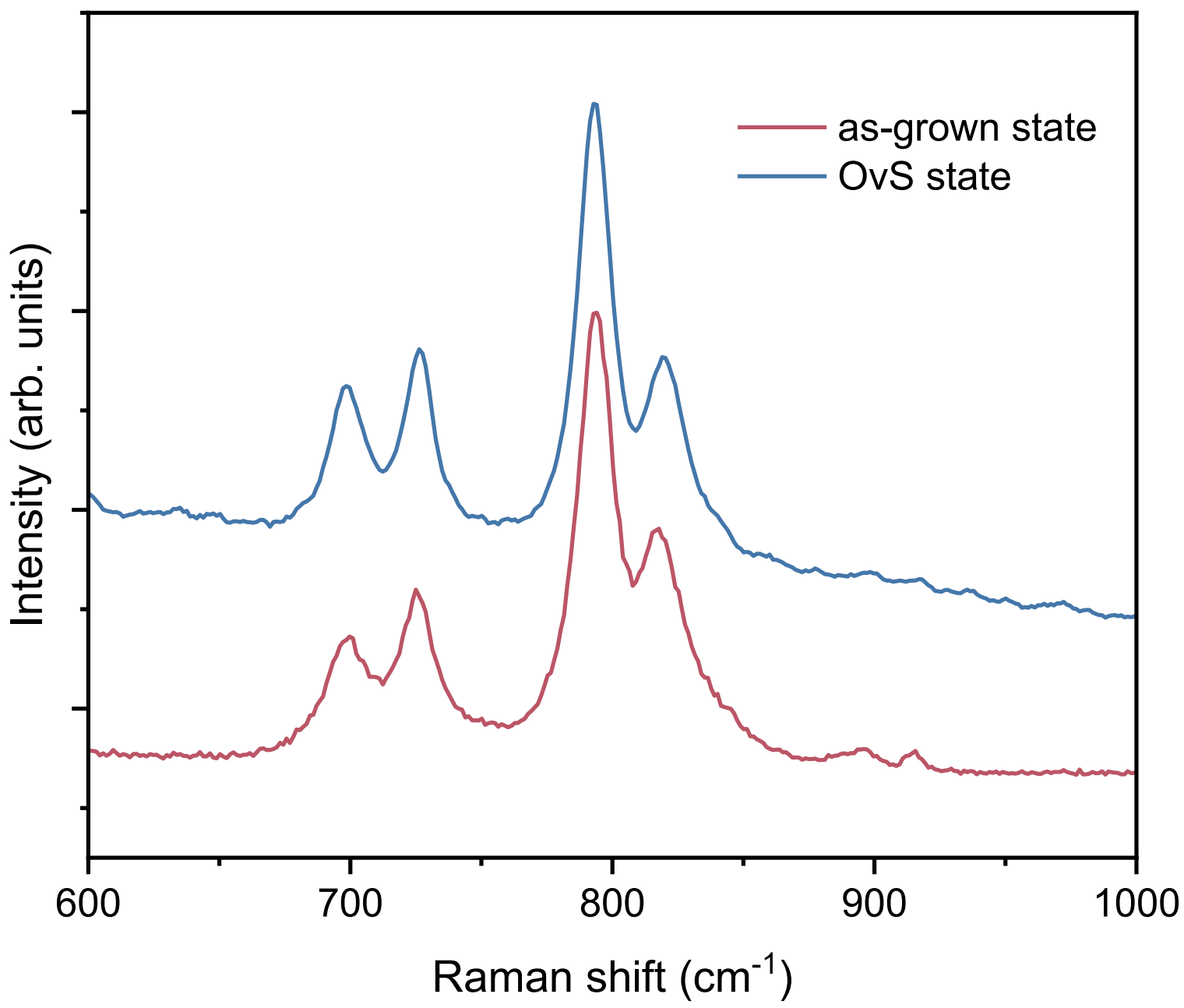

**Figure S4. Unpolarized Raman spectra of the $Bi_2WO_6$ film before and after nitrogen-plasma treatment.**

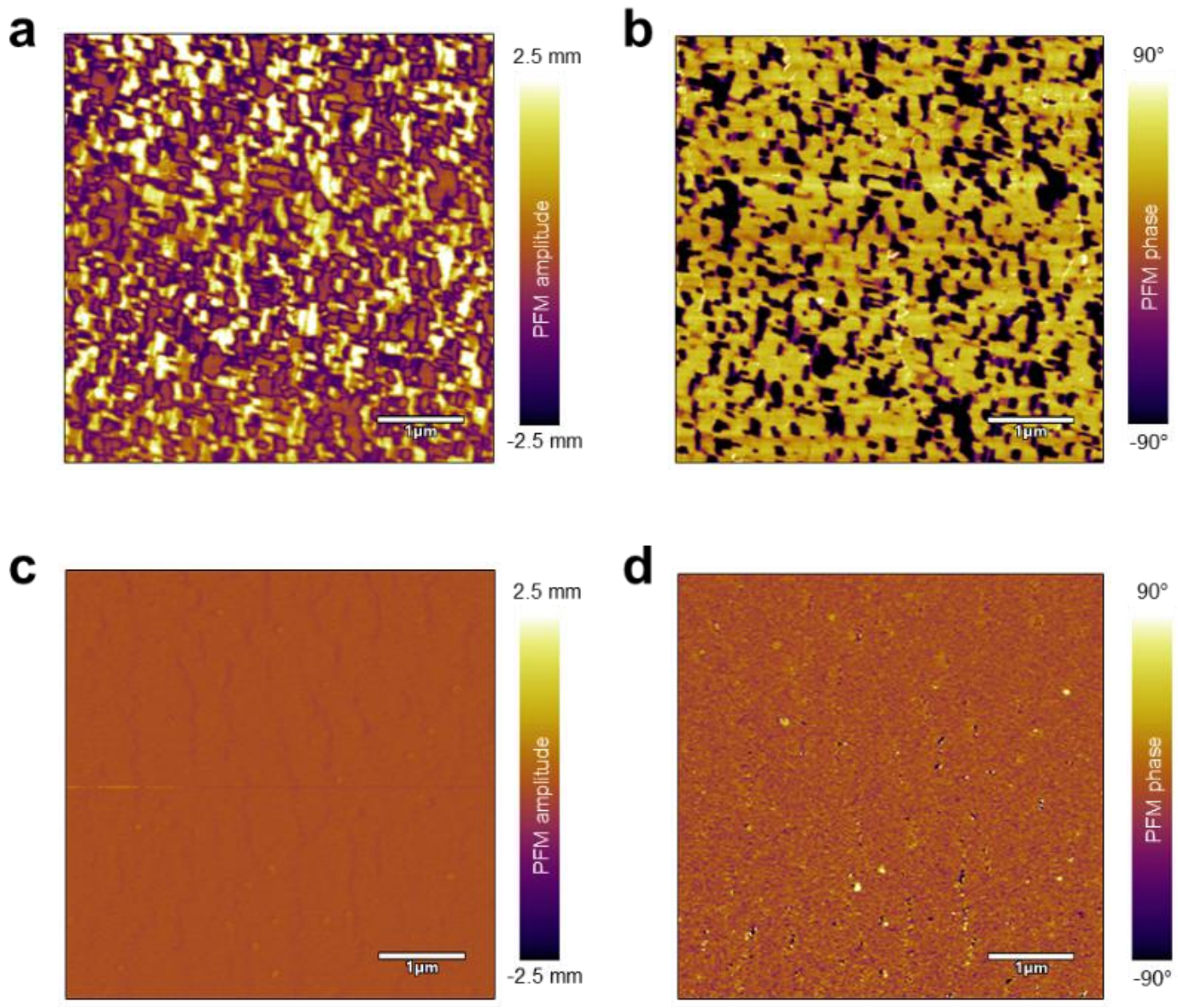

**Figure S5. PFM characterization of $Bi_2WO_6$ films across the colossal resistivity switching** (a, b) PFM amplitude and phase images of the pristine ferroelectric $Bi_2WO_6$ film, showing well-defined ferroelectric domains. (c, d) Corresponding PFM amplitude and phase images of the same film after nitrogen-plasma treatment, revealing the complete disappearance of ferroelectric contrast in the highly

conductive state.

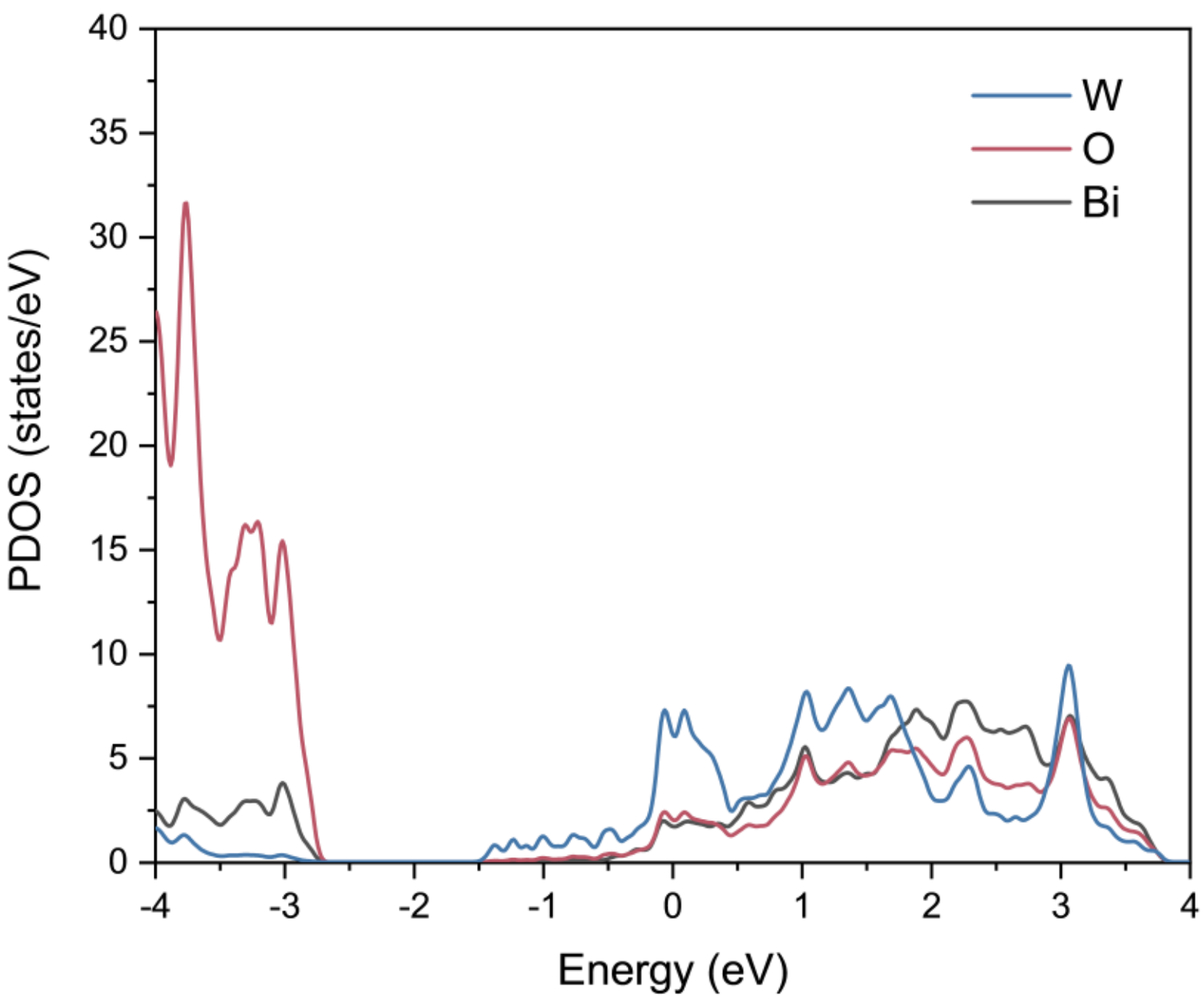

**Figure S6. Partial density of states (PDOS) of the highly conductive OvS state.**

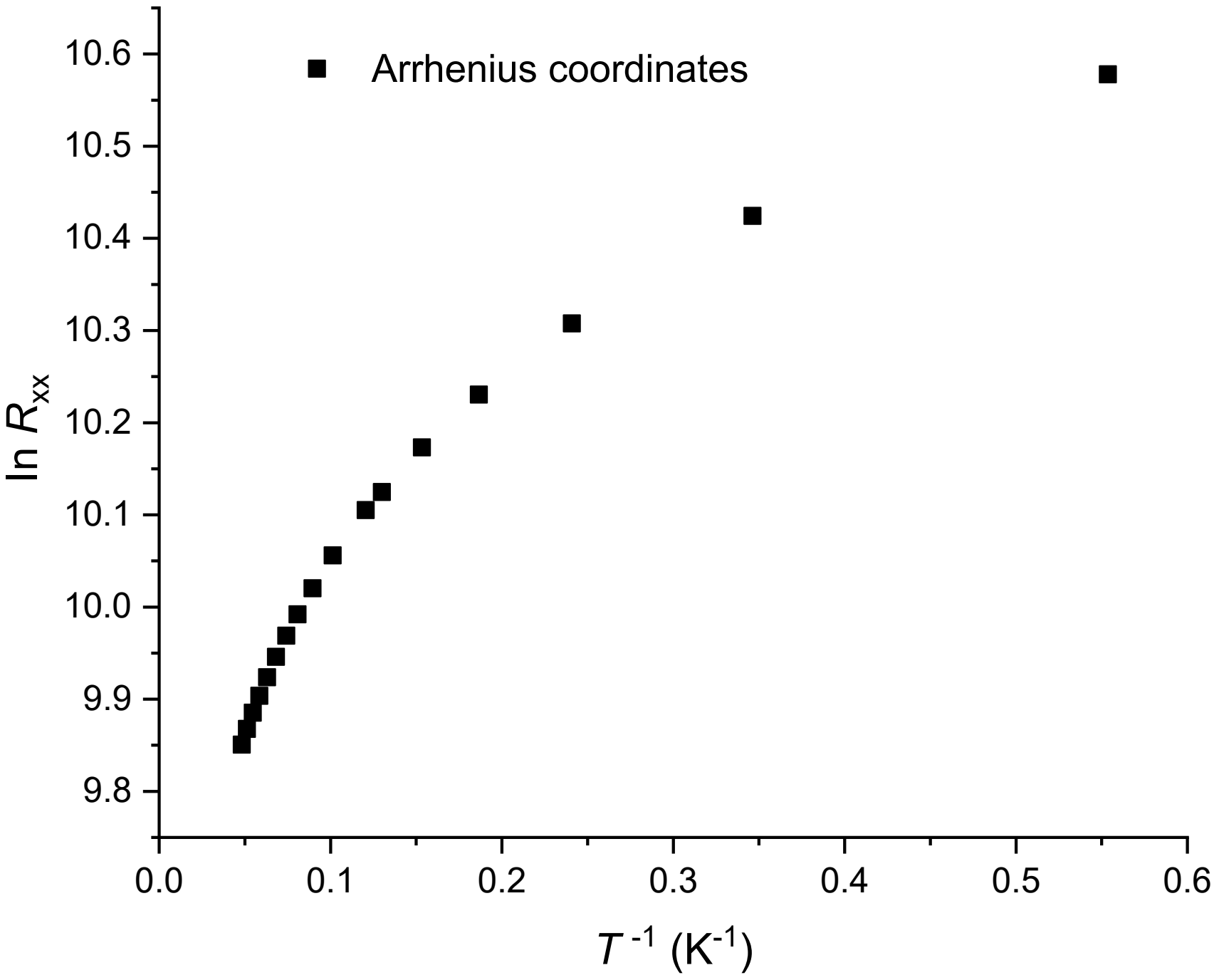

**Figure S7. Arrhenius plot of the low-temperature resistance of the highly conductive state.** The data are plotted as ln$R$ versus $1/T$; the significant deviation from linearity indicates that simple thermal activation cannot describe the transport in this system.

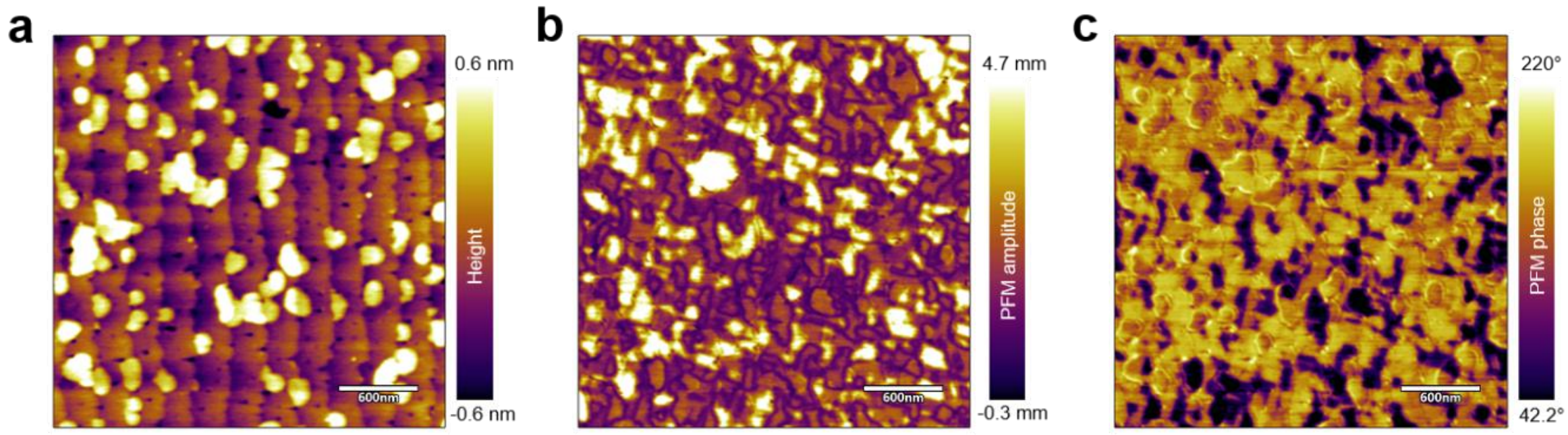

**Figure S8. Topography and PFM characterization of the as-grown 100 nm single-crystalline $Bi_2MoO_6$ films.** (a) Surface topography of the as grown $Bi_2MoO_6$ thin film, showing atomically smooth surface morphology. (b, c) In plane (b) PFM amplitude and (c) phase images of the pristine ferroelectric $Bi_2MoO_6$ film, revealing well-defined ferroelectric domain structures.

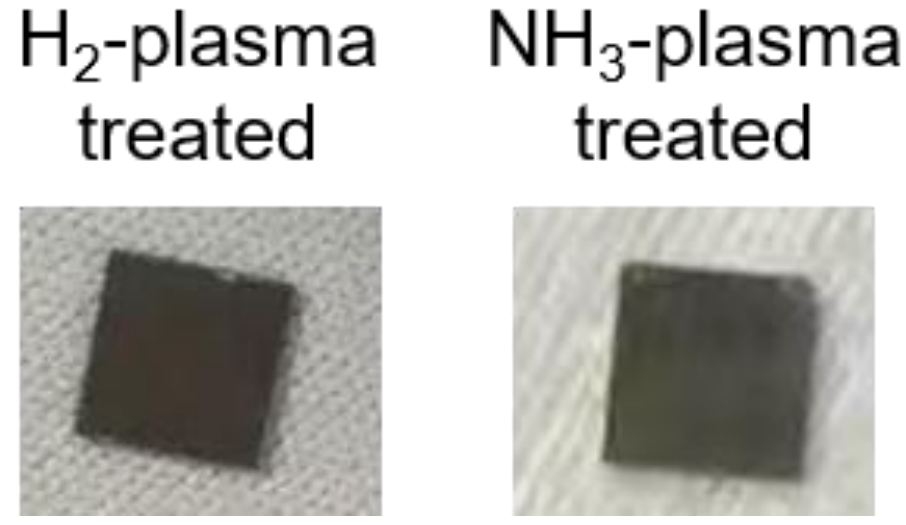

**Figure S9. Optical images of BWO films after $H_2$ (left) and $NH_3$ (right) plasma treatments. Both samples exhibit uniform black coloration, indistinguishable from the $N_2$-plasma-treated film**